\documentclass[prb,twocolumn,superscriptaddress,floatfix]{revtex4}
\usepackage{lmodern} 
\usepackage[T1]{fontenc}
\usepackage{textcomp}
\usepackage[scr=boondoxo,frak=boondox,bb=boondox]{mathalfa}
\usepackage{graphics,epsfig,amsfonts,amssymb,amsmath,color}
\usepackage{subfig}

\begin{document}
\title{Spin-orbit effects surfacing on manganites}
\author{S. Valencia}
\email{sergio.valencia@helmholtz-berlin.de}
\affiliation{Helmholtz-Zentrum Berlin für Materialien und Energie, Albert-Einstein-Str. 15, 12489 Berlin, Germany}
\author{M.J. Calder\'on}
\email{calderon@icmm.csic.es}
\affiliation{Materials Science Factory, Instituto de Ciencia de Materiales de Madrid, ICMM-CSIC, Cantoblanco, E-28049 Madrid, Spain.}
\author{L. L\'opez-Mir}
\affiliation{Institut de Ci\`encia de Materials de Barcelona, ICMAB-CSIC, Campus de la UAB, 08193 Bellaterra, Spain}
\author{Z. Kostantinovic}
\affiliation{Center for Solid State Physics and New Materials, Institute of Physics Belgrade, University of Belgrade, Pregrevica 118, 11080 Belgrade, Serbia}
\author{E. Schierle}
\affiliation{Helmholtz-Zentrum Berlin für Materialien und Energie, Albert-Einstein-Str. 15, 12489 Berlin, Germany}
\author{E. Weschke}
\affiliation{Helmholtz-Zentrum Berlin für Materialien und Energie, Albert-Einstein-Str. 15, 12489 Berlin, Germany}
\author{L. Brey}
\affiliation{Materials Science Factory, Instituto de Ciencia de Materiales de Madrid, ICMM-CSIC, Cantoblanco, E-28049 Madrid, Spain.}
\author{B. Mart\'inez}
\affiliation{Institut de Ci\`encia de Materials de Barcelona, ICMAB-CSIC, Campus de la UAB, 08193 Bellaterra, Spain}
\author{Ll. Balcells}
\affiliation{Institut de Ci\`encia de Materials de Barcelona, ICMAB-CSIC, Campus de la UAB, 08193 Bellaterra, Spain}
\date{\today}
\begin{abstract}
Spin-orbit coupling in magnetic systems lacking inversion symmetry can give rise to non trivial
spin textures. Magnetic thin films and heterostructures are potential candidates for
the formation of skyrmions and other non-collinear spin configurations as inversion symmetry is
inherently lost at their surfaces and interfaces. However, manganites, in spite of their extraordinarily rich
magnetic phase diagram, have not yet been considered of interest within this context as their spin-orbit
coupling is assumed to be negligible. We demonstrate here, by means of angular
dependent X-ray linear dichroism experiments and theoretical calculations, the existence of a noncollinear
antiferromagnetic ordering at the surface of ferromagnetic La$_{2/3}$Sr$_{1/3}$MnO$_3$ thin films
whose properties can only be explained by an unexpectedly large enhancement of the spin-orbit
interaction. Our results reveal that spin-orbit coupling, usually assumed to be very small on
manganites, can be significantly enhanced at surfaces and interfaces adding a new twist to the
possible magnetic orders that can arise in electronically reconstructed systems.
\end{abstract}
\maketitle

Broken inversion symmetry and strong spin-orbit coupling (SOC) are necessary (although not sufficient) conditions for the formation of non-collinear spin textures,~\cite{roszlerNat2006,bodeNature2007,banerjeeNatPhys2013,banerjeePRX2014, liPRL2014,soumyanarayananNat2016,fertNatRevMater2017} such as skyrmions,
magnetic bubbles or spirals. These non-trivial spin configurations, when topological, hold promise for future spin-based information technologies due to their long lifetimes, stability, and the possibility to be created, controlled and/or detected at room temperature by means of electrical currents or electric fields.~\cite{wooNatMat2016,maccarielloNature2018,grossNature2017} While helical and skyrmion states have been observed in several different systems~\cite{muhlbauerScience2009,munzerPRB2010,yuNat2010,fertNatRevMater2017,heinzeNatPhys2011,yuNatMater2011,adamsPRL2012,sekiSci2012,rommingSci2013}  their stability has been found to be enhanced in 2D ones~\cite{banerjeePRX2014,yuNatMater2011,butenkoPRB2010} such as thin films or heterostructures, as space inversion symmetry is inherently broken at surfaces and interfaces. 

Manganese perovskites with formula L$_{1-x}$A$_x$MnO$_3$ (L and A being trivalent lanthanides and divalent alkaline ions, respectively), such as, for example, La$_{2/3}$Sr$_{1/3}$MnO$_3$ (LSMO) have not yet been considered as plausible hosts for non-trivial spin configurations. Although the interplay between spin, lattice, and orbital degrees of freedom leads to a rich phase diagram with a large variety of electronic reconstructions at surfaces and interfaces,~\cite{calderonPRB1999,chakhalianNatPhys2006,breyPRB2007,sefrioui2010,zubkoAnnRevCMP2011,hwangNatMat2012,NgaiAnnRevMR2014,liaoNatMat2016} the atomic spin-orbit coupling in these systems is considered to be negligible. Indeed, it is assumed that the spin-orbit interaction on Mn atoms is rather small $\lambda \sim 0.04$ eV~(Ref.~\citenum{jaoulJMMM1977}) compared to the typical energy scale on manganites given by the hopping parameter $t\sim 0.2-0.5$ eV. Moreover, the e$_g$ orbitals of Mn ($x^2-y^2$ and $3z^2-r^2$), that host the conduction electrons in manganites, are not directly spin-orbit coupled by symmetry.~\cite{malozemoff1986} The coupling arises only through the t$_{2g}$ orbitals ($xy$, $yz$, $zx$) as a second order process.~\cite{fuhrJPCM2010} Hence the SOC between the e$_g$ orbitals is given by $g=\lambda^2/\Delta$, being $\Delta\sim 1.5$ eV the crystal field splitting between the e$_g$ and t$_{2g}$ levels, being one order of magnitude smaller than $\lambda$. Such small SOC is consistent with accordingly small anisotropic responses observed in bulk.~\cite{calderonPRB2001,fuhrJPCM2010,nemesAdvMat2014}

Here we show that a non-negligible SOC, as such required for the appearance of non-collinear spin textures, arises on the surface of manganite thin films. We make use of X-ray linear dichroism (XLD) to characterise an optimally doped (x=1/3) LSMO thin film. Our results confirm previous ones reporting the existente of a surface antiferromagnetic (AFM) layer~\cite{arutaPRB2009,valenciaJPCM2014,pesqueraPRA2016} on an otherwise ferromagnetic (FM) compound. However, in contradiction with previous conjectures, we demonstrate that the AFM axis of the surface in the presence of a magnetic field is not necessarily parallel or perpendicular to the surface. By means of angular dependent XLD measurements we show that the relative orientation of the AFM axis with respect to the bulk FM ordering axis does depend on the angle at which the magnetic field is applied with respect to the film plane.  We demonstrate that this result can only be explained by the enhancement of the SOC on the surface by at least one order of magnitude. Such enhancement, together with the inherent breaking of the spatial inversion symmetry at the surface of the films, makes manganites a plausible candidate for the creation and control of non-trivial spin configurations. 

\begin{figure*}
\leavevmode
\includegraphics[clip,width=0.8\textwidth]{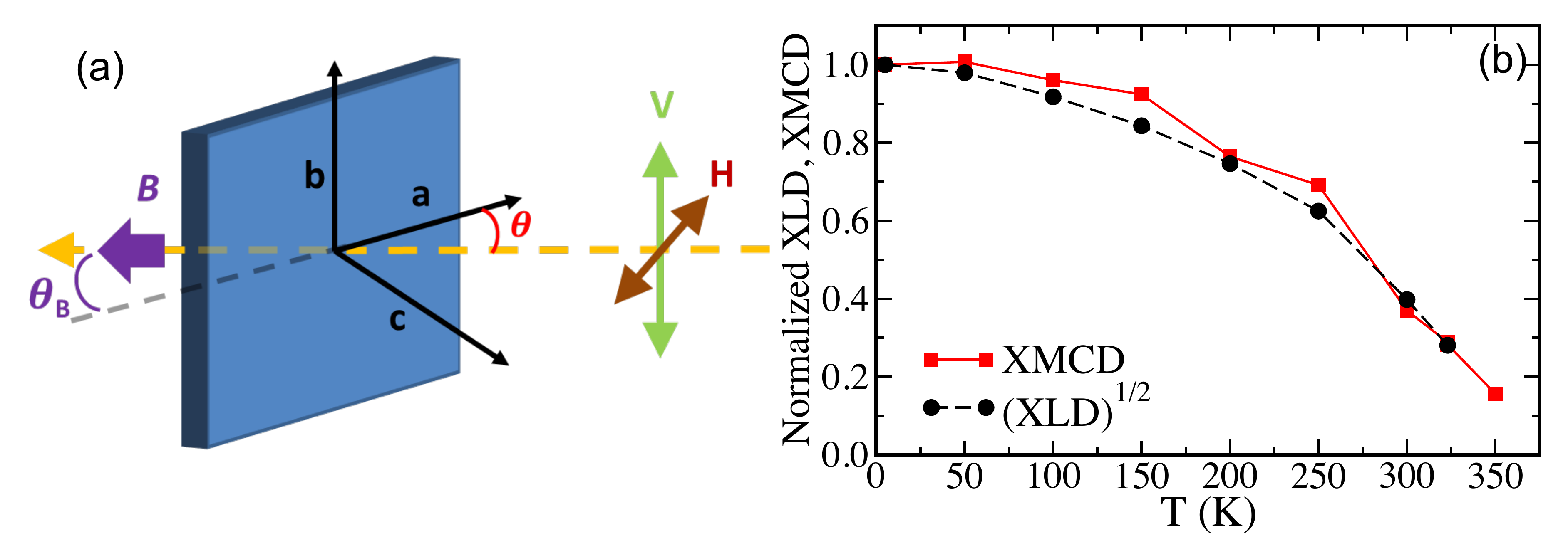}
\caption{a) Sketch of the experimental setup and geometry used for the XLD measurements. b) Comparison of the temperature dependence of the normalized XMCD and XLD measured at the Mn $L_3$ and $L_2$ edges, respectively (see Appendix A). The XLD data is square rooted to account for its square dependence on magnetisation as opposed to the XMCD linear dependence. Both curves show the same $T$ dependence confirming the AFM origin of the XLD signal (see text for discussion).}
\label{fig:setup-XMCD-XLD}
\end{figure*}

High quality LSMO thin films with thickness ranging between $1.5$ nm and $45$ nm were deposited on $(001)$-oriented SrTiO$_3$ (STO) substrates by means of magnetron sputtering.~\cite{konstantinovicJAP1999} The experimental characterization of these samples (Refs.~\citenum{valenciaJPCM2014},~\citenum{sandiumengePRL2013}  and Appendix A) shows thickness dependent results in agreement with previous reports in terms of magnetic and electric properties as well as in terms of preferential orbital occupation.~\cite{TebanoPRL2008,PesqueraNatComm2012,cuiSciRep2014, arutaPRB2009,valenciaJPCM2014,pesqueraPRA2016}
In particular, the $9.4$ nm thick LSMO film under study exhibits FM and metallic behaviour with magnetic transition temperature T$_C$ and magnetisation values very close to those of the bulk~\cite{sandiumengePRL2013} and a preferential $x^2-y^2$  orbital occupation that corresponds to the tensile strain conditions imposed by the STO substrate.~\cite{TebanoPRL2008,baenaPRB2011,PesqueraNatComm2012,cuiSciRep2014}

The characterization of surface layers, in terms of oxidation states, orbital occupancy and magnetic properties, is possible thanks to element-selective synchrotron-related techniques.~\cite{luningPRB2003,arutaPRB2009,liuJPCM2015} In particular, XLD, calculated from the difference between two absorption spectra ($I^V$ and $I^H$) taken for incoming vertical (V) and horizontal (H) linearly polarized radiation (Figure~\ref{fig:setup-XMCD-XLD}(a)), probes the anisotropies around the element under investigation. Such anisotropies can arise due to a preferential orbital occupancy (XLD$_{\rm OO}$) and/or by the existence of ferromagnetic (XLD$_{\rm FM}$) and antiferromagnetic (XLD$_{\rm AFM}$) orderings along specific crystallographic directions. 

In XLD measurements, the electric field vector $\bf{E}$ of the incoming radiation probes different sample directions, see Fig.~\ref{fig:setup-XMCD-XLD}(a). Axis parallel to (100)$_{\rm LSMO}$, (010)$_{\rm LSMO}$ and (001)$_{\rm LSMO}$ crystallographic directions have been labelled $a$, $b$ and $c$, respectively. Within our experimental geometry, a vertically polarized incoming beam corresponds to $\bf{E}$ parallel to $b$ independently of the angle of incidence $\theta$, i.e. $I_\theta^V=I_b$. On the other hand, the electric field vector for incoming horizontally polarized photons has components along $a$ and $c$ directions, with their relative weights depending on $\theta$.

In order to obtain accurate details of the surface magnetic configuration and, more concretely, about the relative orientation of the FM-bulk and AFM-surface axis, we have performed XLD measurements at different angles of incidence on a 9.4 nm thick LSMO thin film. A magnetic field of 3 T aligns the LSMO magnetisation along the beam propagation direction for all $\theta$. Within these conditions, the FM axis is always orthogonal to the electric field vector of both V and H polarised beams such that the FM order does not contribute to the XLD measurement (Ref.~\citenum{arutaPRB2009} and Appendix A). The in-plane symmetry of LSMO implies that $I_a$ = $I_b$ = $I_{ab}$ such that XLD$_{\theta}$ = $I_\theta^V-I_\theta^H$ = $I_{ab}-I_\theta^H$.

Fig.~\ref{fig:setup-XMCD-XLD}(b) shows the temperature dependence of XLD$_{30^o}$. Given that XLD$_{\rm FM}$ is suppressed and XLD$_{\rm OO} \ll {\rm XLD}_{\rm AFM}$ (see Appendix A) the temperature dependence of the XLD reflects that of the surface related~\cite{arutaPRB2009,valenciaJPCM2014,pesqueraPRA2016} AFM phase. We also depict in Fig.~\ref{fig:setup-XMCD-XLD}(b) the temperature dependence of the X-ray magnetic circular dichroism (XMCD), see Appendix A. XMCD, being linearly dependent on the magnetisation, is only sensitive to the FM bulk component of the film.~\cite{carraPRL1993} The close resemblance between both curves corroborates the antiferromagnetic origin of the obtained XLD signal. 

\begin{figure}
\leavevmode
\includegraphics[clip,width=0.42\textwidth]{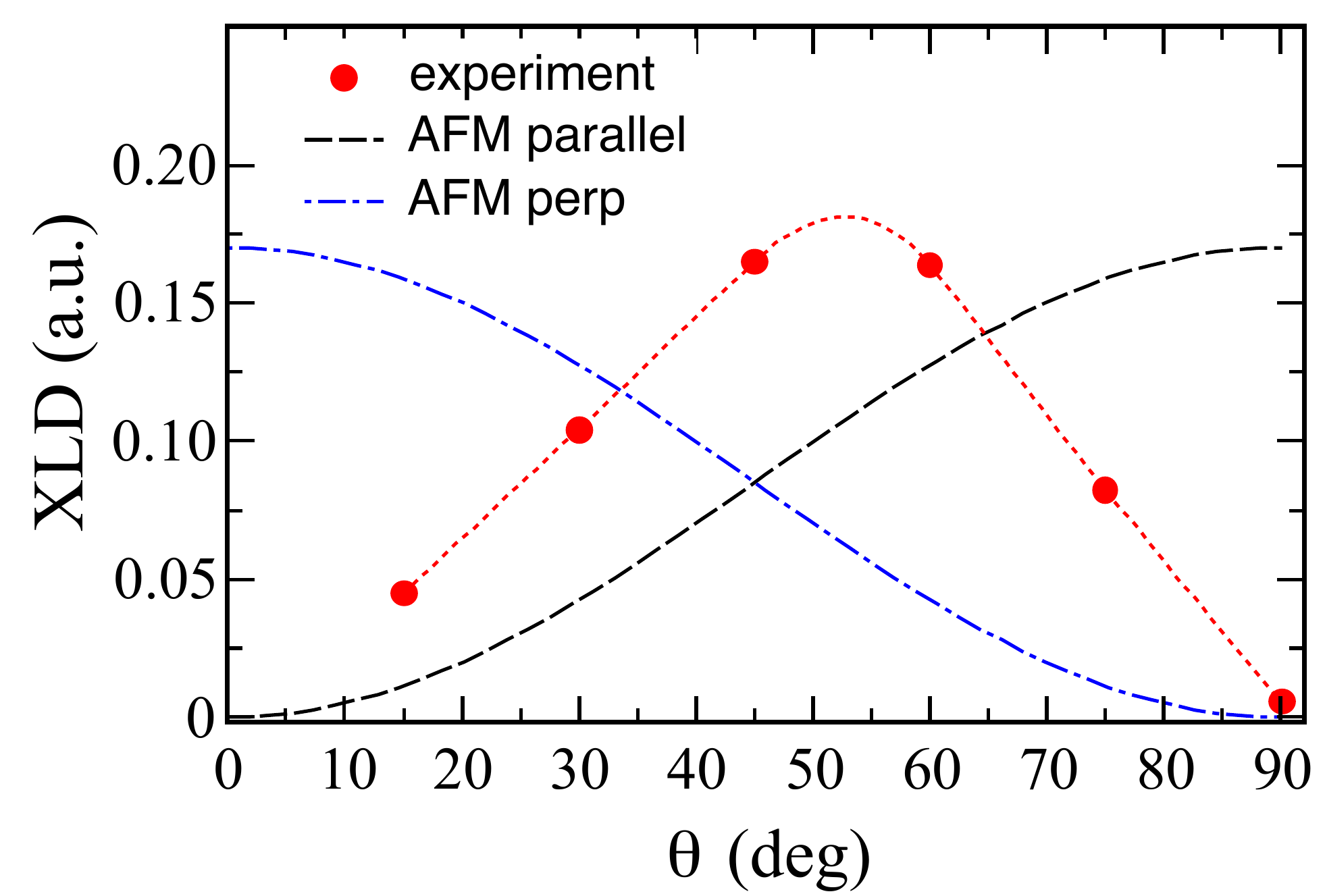}
\caption{Experimentally measured angular dependence of the XLD (red dots) at the Mn $L_2$ edge for a thin (9.4 nm) LSMO thin film. The dotted line is a guide to the eye. For comparison we show the expected XLD angular dependent behaviour for the case of AFM order with magnetic axis oriented parallel (dashed black) or orthogonal (dashed-dotted blue) to the surface plane of the sample.}
\label{fig:XLD-exp}
\end{figure} 

\begin{figure*}
\leavevmode
\includegraphics[clip,width=0.8\textwidth]{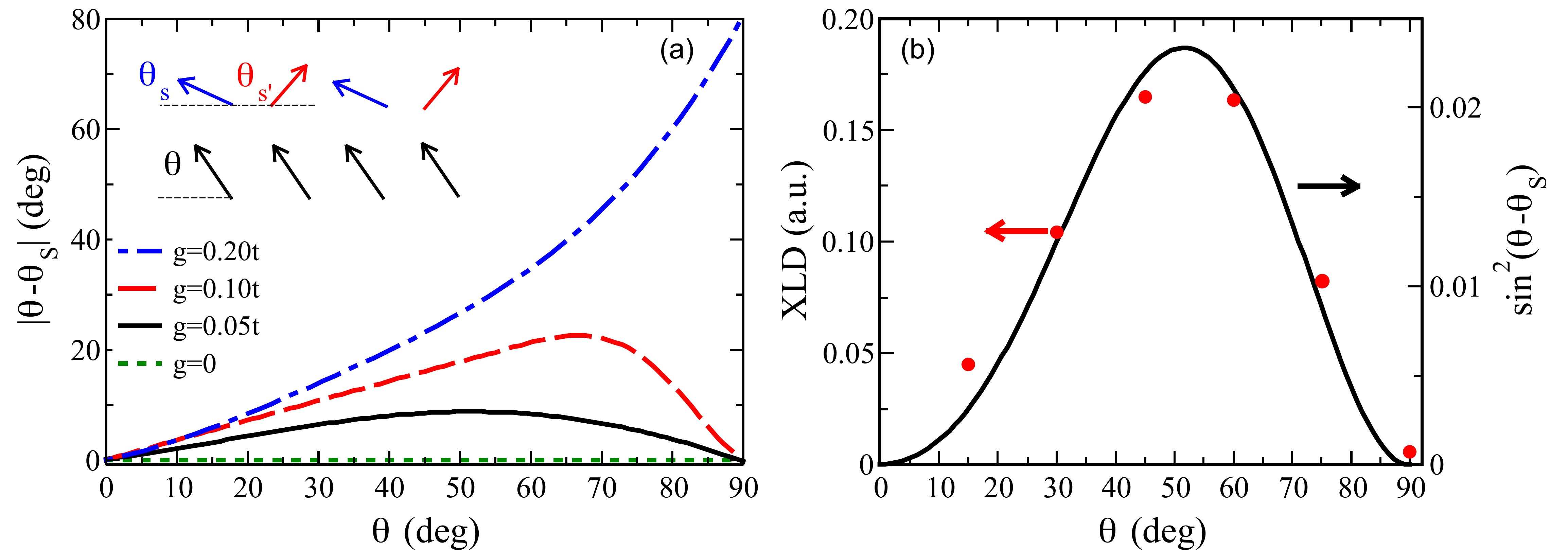}
\caption{a) Predicted misalignment between the FM and AFM axis  $|\theta-\theta_s|$ vs $\theta$ for different values of the interfacial SOC $g$. Inset: Sketch of the spins showing the angle between the bulk FM ($\theta$) and surface AFM ($\theta_s,\theta_s'$) axis with respect to the sample's surface. The ground state for the surface gives $\theta_s'=\theta_s+\pi$. b) Comparison of the experimentally measured (red dots) and the predicted (solid line) XLD angular dependence for $g = 0.05 t$.}
\label{fig:theory}
\end{figure*} 

Angular dependent XLD experiments have been done at the lowest possible temperature (4 K) in order to enhance the AFM contribution to the dichroic signal. As in the temperature dependence case, a magnetic field of 3 T has been applied along the beam propagation direction in order to suppress the FM component. Under these conditions it can be shown~\cite{luningPRB2003} that  XLD$_\theta \propto \cos^2 \phi$ where $\phi$ is the angle between {\bf E} and the AFM axis.

Based on XLD measurements restricted to one or two values of $\theta$, the existence of an AFM surface layer with magnetisation axis orthogonal to the sample surface on LSMO has been previously reported.~\cite{valenciaJPCM2014,arutaPRB2009,pesqueraPRA2016} If this were the case, XLD$_{\theta} $ would be a 180$^o$ periodic curve as that depicted in Fig.~\ref{fig:XLD-exp} (blue dot-dashed line) with maximum XLD at $\theta=0^o$. Likewise, an in-plane AFM orientation would show identical periodicity but shifted by 90$^o$ (black dashed line). An intermediate case where the AFM axis is tilted by an angle $\gamma$ with respect to the sample's normal would lead to a curve shifted by $\gamma$. The experimentally measured angular dependence of the XLD shows none of these angular dependences (red dots on Fig.~\ref{fig:XLD-exp}). Indeed, we find that XLD$_{\rm AFM}$ increases from XLD $\sim  0$ at gracing incidence ($\theta = 0^o$) towards a maximum at $45^o < \theta < 60^o$ followed by a decrease back towards XLD $\sim 0$ for ($\theta = 90^o$). This behavior also excludes a scenario where the angle defined by the AFM and FM axis is the same for all $\theta$ as in that case we would expect  XLD$_{\theta}$ to be constant.  
Hence we conclude that the relative orientations between (i) the AFM and the FM axis and (ii) the AFM axis and the surface plane of the sample depend on the angle at which the magnetic field is applied. Namely, the AFM axis is collinear to the FM one (XLD$\sim 0$) only for $\theta=0^0$ (FM and AFM in-plane) and for $\theta=90^0$ (FM and AFM fully out-of-plane). For intermediate angles, the FM and AFM axis are not collinear (XLD $\ne 0$), have and out-of-plane component, and their relative orientation depends on $\theta$. 

The experimental data in Fig.~\ref{fig:XLD-exp} cannot be explained within the standard model for manganites involving FM double exchange, AFM superexchange interactions and Jahn-Teller distortions as within this model the AFM ordering axis is always collinear with the bulk FM one, i.e. XLD$_{\theta}$ would be constant. A canting of the AFM axis due to the 
Dzyaloshinskii-Moriya interaction (DMI)~\cite{dzyaloshinskiiJPCS1958,moriyaPR1960} derived as a relativistic correction to superexchange~\cite{moriyaPR1960} ${\bf D_{ij}} \cdot ({\bf S}_i \times {\bf S}_j)$ is not plausible either. Note that the inversion symmetry breaking with respect to the $x-y$ surface would involve a vector coupling $(0,0,D_z)$ which cannot give rise to a canting of the magnetic order axis in the $z$ direction, as observed in the experiment. 
 
To find the physical origin which explains the observed angular dependende of the XLD we consider a model which incorporates (i) the kinetic energy through the double exchange interaction~\cite{zenerPR1951} which includes the two $e_g$ orbitals~\cite{calderonPRB1999} with hoppings given by the Slater Koster parametrization,~\cite{slaterPR1954} (ii) the long range Coulomb interaction between charges in the system which affects the redistribution of charge close to the surface,~\cite{breyPRB2007} and (iii) the SOC between the e$_g$ orbitals (Ref.~\citenum{fuhrJPCM2010} and Appendix B). We checked that Jahn-Teller coupling, which is expected not to be important in La$_{2/3}$Sr$_{1/3}$MnO$_3$, cannot explain the observations and was therefore disregarded.
We also neglected the lattice buckling on the surface~\cite{prunedaPRL2007} as its effect on the hoppings~\cite{sergienkoPRL2006,slaterPR1954} is estimated to be much smaller than the effect of the broken inversion symmetry. In the model we use a one-atomic-plane thick surface, with a total of twelve atomic layers and periodic boundary conditions.
All energies are given in units of $t$, the hopping parameter. The hopping processes are orbital dependent and anisotropic, with $x^2-y^2$ producing a 2D band in the $ab$ plane and $3z^2-r^2$ having a larger hopping along the $c$-direction. The SOC between the e$_g$ orbitals occurs as a second order process which involves the t$_{2g}$ orbitals~\cite{fuhrJPCM2010}

\begin{equation}
H_{\rm SOC}=g \left( \begin{array}{ccc}
 3 \cos^2(\theta_{mag})  &  \sqrt{3} \cos^2(\theta_{mag})   \\
\sqrt{3} \cos^2(\theta_{mag})  &  \cos^2(\theta_{mag})+4 \sin^2(\theta_{mag})
 \end{array} \right)
 \label{eq:SOI}
\end{equation}
with $\theta_{mag}$ being the magnetic moment angle with respect to the surface plane and $g$ the second order spin-orbit coupling. We only consider H$_{\rm SOC}$ at the surface. By itself, H$_{\rm SOC}$ adds a small shift ($\propto g$) to the onsite energies in such a way that, at the surface, the $3z^2-r^2$ orbital is shifted down for $\theta_{mag}= 0$ while $x^2-y^2$ is shifted down for $\theta_{mag}= \pi/2$. The bulk FM ordering axis $\theta_{mag}$ is fixed by a large magnetic field (3 T in the experiment) applied along the beam propagation direction $\theta$ (see Fig.~\ref{fig:setup-XMCD-XLD}(a)) such that $\theta_{\rm mag}=\theta$.
For modelling the film surface, its spins are considered to be in two interpenetrating sublattices, such that all first nearest neighbors of one sublattice are on the complementary one, and allow the spins to rotate with angles $\theta_s$ and $\theta_s'$, respectively, see inset of Fig.~\ref{fig:theory}(a). The spin configuration with the lowest energy corresponds to $\theta_s'=\theta_s+\pi$, i.e. an AFM surface. 
The AFM order at the surface comes about because the boundary conditions on the surface,~\cite{breyPRB2007,salafrancaPRB2008,calderonPRB2008} suppressing the FM interaction generated by the double exchange mechanism.~\cite{calderonPRB1999} Moreover, in a (001) surface, the $x^2-y^2$ orbital is not connected to the bulk (the hopping in the $c$-direction is zero) and hence the $3z^2-r^2$ orbital is preferentially occupied. 

We calculate the energy of the system as a function of the orientations of the FM bulk $\theta$ and the AFM surface for different values of the SOC $g$. 
The resulting non-collinearity or misalignment between both angles $|\theta-\theta_s|$ is plotted as a function of $\theta$ in Fig.~\ref{fig:theory}(a).

Our theoretical results show that for $g=0$ the bulk FM ordering and the surface induced AFM ordering are collinear, namely $\theta=\theta_s$, for all values of $\theta$. 
For $0.05 t  \leq g \leq 0.1 t$, $|\theta-\theta_s|$  varies continuously as a function of $\theta$ with a maximum at an intermediate angle between 0$^o$ and 90$^o$ which depends on $g$. Hence, for $0.05 t  \leq g \leq 0.1 t$, the spins on the surface still rotate but the AFM ordering axis is not collinear with the bulk FM one and the relative angle is a function of $\theta$. This relative tilting of the AFM ordering axis occurs due to the subtle competition between the collinear axis favored by the double exchange plus Coulomb terms and the small relative angle favored by the enhanced SOC. For even larger $g$, the AFM ordering axis on the surface is stuck to a small angle ($\theta_s$) with respect to the surface plane in order to keep on favoring the occupation of the $3z^2-r^2$ orbital.

An AFM ordering axis tilted by  $\theta_s$ with respect to the sample surface leads to a linear dichroic signal XLD$_{\theta}$ = $I_{ab}-I_\theta^H \propto  \cos^2 \phi= \sin^2 (\theta-\theta_s$). Fig.~\ref{fig:theory}(b) shows the predicted $\sin^2 (\theta-\theta_s)$ vs $\theta$ dependence for the $g = 0.05t$ case (solid curve) together with the experimentally measured angular dependence of XLD (dots) showing an excellent qualitative agreement. The values of $g$ required to reproduce the experimental results amount (considering $t \sim 0.2-0.5$ eV) to $\sim 0.01-0.025$ eV. These values are significantly larger than the estimate $g=\lambda^2/\Delta=0.001$ eV. An increase of $g$ by at least one order of magnitude can not be solely explained by the decrease of the t$_{2g}$-e$_g$ splitting $\Delta$ that may take place at the interface.~\cite{burtonPRB2016} The SOC effective enhancement is likely due to a Rashba spin-orbit contribution produced by the electric field caused by the redistribution of charge near the interface (electronic reconstruction).~\cite{breyPRB2007,salafrancaPRB2008,calderonPRB2008}

Spin-orbit coupling is behind many of the effects observed in manganites like the anomalous Hall effect~\cite{calderonPRB2001} or anisotropic magnetoresistance.~\cite{fuhrJPCM2010,nemesAdvMat2014} However, in those cases the observations could be explained with the usually assumed small value for SOC. Our observation of an unusually large SOC implies that a Rashba-like DMI~\cite{banerjeePRX2014,kunduPRB2015} could be present giving rise to complex spin textures like skyrmions and spirals on manganite surfaces. The large SOC 
could also explain the large tunnelling anisotropic magnetoresistance measured in a related device.~\cite{galceranAIPAdv2016} Similar ideas have been proposed for the magnetic metallic gas formed at interfaces in LaAlO$_3$/SrTiO$_3$ heterostructures, a system in which Rashba spin-orbit coupling can be tuned,~\cite{caviglia2010} and the magnetic order has been argued to be a long wave-length spiral.~\cite{banerjeeNatPhys2013} The case of manganite heterostructures is potentially more interesting due to the variability of magnetic orders they can sustain.

In summary, we have performed x-ray linear dichroism measurements as a function of the incidence angle that reveal an antiferromagnetic order at the surface of a ferromagnetic and metallic La$_{2/3}$Sr$_{1/3}$MnO$_3$ thin film. The antiferromagnetic order axis is non-collinear to the ferromagnetic bulk one. This result can only be explained by introducing a significant spin-orbit coupling, much larger than previously assumed for manganites. This large spin-orbit coupling implies that manganite surfaces or interfaces, where the inversion symmetry is broken, might constitute a new scenario for the appearance of non-trivial spin textures, opening, in this way, a new avenue for exploration and applications of manganite heterostructures. 

{\it Acknowledgement.} The experimental research leading to these results has received funding from the European Community's Seventh Framework Programme (FP7/2007-2013) under grant agreement No 312284.  MJC and LB acknowledge funding from Ministerio de Econom\'ia, Industria y Competitividad MEIC (Spain) via Grant No FIS2015-64654-P. ZK thanks Project III45018 from the Ministry of Education and Science of Republic of Serbia. LLM, BM and LlB thanks MEIC through the Severo Ochoa Programme for Centres of Excellence in R\&D (SEV-2015-0496) and MAT2015-71664-R.

SV and MJC have contributed equally to this work. 

\renewcommand{\thefigure}{A\arabic{figure}}
\renewcommand{\theequation}{A\arabic{equation}}
\setcounter{figure}{0} 
\setcounter{equation}{0} 
\appendix
\begin{figure*}
\leavevmode
\includegraphics[clip,width=0.8\textwidth]{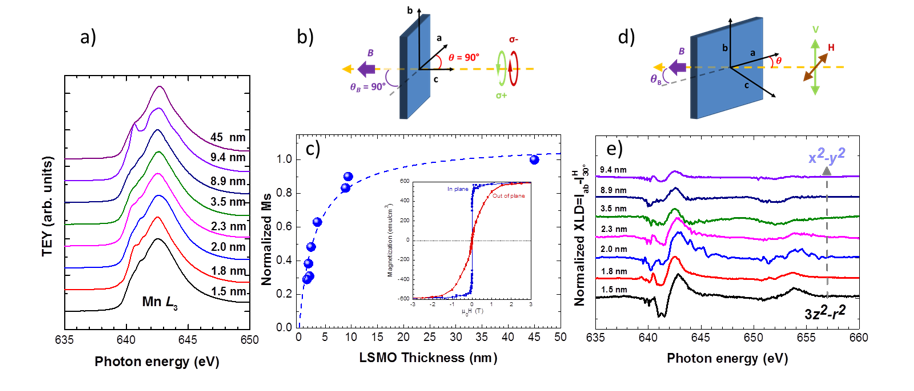}
\caption{a)  Mn $L_3$-edge XAS vs LSMO thin film thickness. The $45$ nm sample shows bulk like spectral shape. Decreasing the film thickness leads to an enhancement of the spectral weight at the low energy side of the $L_3$ spectral feature related to the increase of a Mn$^{3+}$ component. The $9.4$ nm sample shows a clear peak at ca. $641$ eV related to the presence of Mn$^{2+}$. b) Sketch of the experimental setup used for the XMCD measurements. c) Main panel: Normalized spin contribution to the magnetic moment of Mn vs thickness as deduced from the application of the sum rules to the XMCD spectra obtained at normal incidence at $5$ K and H = $3$T. Data has been normalized to the magnetisation obtained for the thicker bulk-like film. Inset: Magnetic hystheresis loops obtained at T = $10$ K for magnetic fields applied parallel and orthogonal to the sample plane. d) Sketch of the experimental setup used for the XLD measurements. e) Orbital contribution to the XLD obtained at $350$ K. At low thicknesses the sign and shape of the XLD at the Mn $L_2$-edge indicates a preferential 3z$^2$-r$^2$ orbital occupancy. Increasing the thickness leads to the observation of an XLD compatible with a preferential x$^2$-y$^2$ orbital occupation.}
\label{fig:appendix1}
\end{figure*} 

\section{Sample characterization}
High quality La$_{2/3}$Sr$_{1/3}$MnO$_3$ thin films with thickness ranging between $1.5$ nm and $45$ nm have been deposited on top of (001)-oriented SrTiO$_3$ (STO) substrates by means of magnetron sputtering.~\cite{konstantinovicJAP1999} The spectroscopic characterization of these samples (below) shows results in agreement with previous reports; i) depressed magnetic properties as film thickness is reduced,~\cite{sandiumengePRL2013} ii) a progressive change from a preferential $3z^2-r^2$  orbital occupation for thinner ($t_{\rm LSMO}  \leq 2.3$ nm) to a $x^2-y^2$ one for thicker ($t_{\rm LSMO}  \geq 3.5$ nm) films,~\cite{TebanoPRL2008,PesqueraNatComm2012,cuiSciRep2014} iii) vanishingly small orbital contribution to the XLD at temperatures below the magnetic transition temperature (i.e. XLD $\simeq$ XLD$_{\rm AFM}$),~\cite{arutaPRB2009} and iv) robust XLD$_{\rm AFM}$ component not depending on deviations of the nominal Mn$^{3+}$/Mn$^{4+}$ composition.~\cite{valenciaJPCM2014,PesqueraNatComm2012,pesqueraPRA2016,jongPRB2006} In this manuscript, we focus on the $9.4$ nm thick sample, with magnetic and electric properties very similar to the bulk ones.

{\it X-ray spectroscopic measurements.}

XAS (X-ray absorption spectroscopy), XMCD (X-ray magnetic circular dichroism) and XLD (x-ray linear dichroism) have been measured across the Mn $L_{2,3}$ edges.  Surface sensitivity is gained when the data is acquired by using the total electron yield (TEY) mode.
The XMCD, proportional to the projection of the magnetisation along the beam propagation direction, is obtained by calculating the difference between the spectrum measured with incoming right ($\sigma^+$) and left ($\sigma^-$) helicity circularly polarized beams and is defined as XMCD = $\sigma^+$-$\sigma^-$, see Fig.~\ref{fig:appendix1}(b). The data have been acquired at normal incidence ($\theta = 90^o$) and with a 3 T magnetic field applied along the beam propagation direction. The XAS spectra have been computed from XAS = $\sigma^+$+$\sigma^-$.

XLD spectra is defined as the difference in absorption between vertical (V) and Horizontal (H) polarized radiation at a gracing incidence angle $\theta$ i.e. XLD$_\theta$=$I_\theta^V-I_\theta^H$, see Fig.~\ref{fig:appendix1}(d). The angle of incidence of the beam with respect to the sample surface has been changed by rotating the sample along its b axis (see Fig.~\ref{fig:appendix1}(b)). The relative weight of the orbital (XLD$_{\rm OO}$), ferromagnetic (XLD$_{\rm FM}$) and antiferromagnetic (XLD$_{\rm AFM}$) contributions to the total XLD signal depends on temperature and on the applied magnetic field. At temperatures above the Curie one (T$_C$) the only contribution to the XLD, as long as the AFM ordering has a critical temperature below T$_C$, is due to the orbital occupancy i.e. XLD = XLD$_{\rm OO}$. At temperatures below T$_C$, both FM and AFM contributions are also present, i.e. XLD = XLD$_{\rm OO}$ + XLD$_{\rm FM}$+XLD$_{\rm AFM}$. A magnetic field of 3 T, strong enough to saturate the magnetisation of the sample along the beam propagation direction, has been applied during XLD acquisition so that the XLD$_{\rm FM}$ component can be cancelled. This cancellation takes place because within these conditions the FM axis is forced to be along the beam propagation direction independently of $\theta$, hence being orthogonal to the electric field vector of both V and H polarized beams. In that case XLD = XLD$_{\rm OO}$ +XLD$_{\rm AFM}$.

\begin{figure}
\leavevmode
\includegraphics[clip,width=0.45\textwidth]{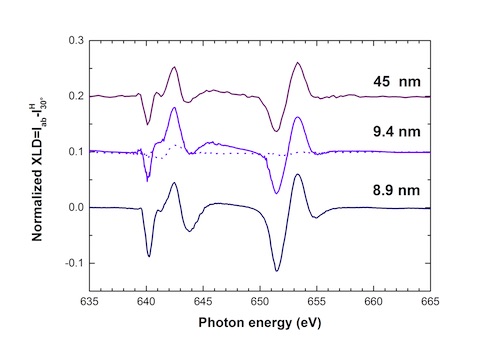}
\caption{XLD spectra obtained at 4 K with a magnetic field applied perpendicularly to the propagation direction of the incoming beam for samples with thickness 8.9 nm, 9.4 nm and 45 nm (continuous lines). For comparison we show the orbital contribution for the 9.4 nm sample (dotted line). The low temperature XLD is similar in all cases and does not depend on the presence of other Mn oxidation states besides that of the mixed Mn$^{3+}$/Mn$^{4+}$ valence one.  }
\label{fig:appendix2}
\end{figure} 

{\it Spectroscopic results.}

{\it Mn valence}. Compared to the other thin films studied by our group, the XAS measurement for the $9.4$nm thick sample presents a shoulder at ca. $641$ eV, see Fig.~\ref{fig:appendix1}(a), which is most likely due to the presence of Mn$^{2+}$ (Refs.~\onlinecite{valenciaJPCM2014,jongPRB2005,valenciaPRB2006,valenciaPRB2007}). Previous results have shown that Mn$^{2+}$ can originate as resulting of a deoxygenation process related to preexisting structural defects~\cite{jongPRB2006,jongPRB2005,valenciaPRB2006,valenciaPRB2007} which might lead to a time dependent formation of divalent Mn.~\cite{valenciaPRB2006} Indeed the $9.4$ nm sample is the oldest of the whole set. Importantly, previous resonant photoemission experiments have shown the strong localization of Mn$^{2+}$ and have excluded its interaction with the major Mn mixed valence Mn$^{3+}$/Mn$^{4+}$ phase.~\cite{jongPRB2005} Moreover, recent results have shown that the spectral shape of the XLD, both at T $>$ T$_C$ and T$ < $T$_C$, is not affected by the presence of Mn$^{2+}$ [\onlinecite{pesqueraPRA2016}]. Indeed, our XLD spectra for $45$, $9.4$ and $8.9$ nm thick films have the same overall shape, see Fig.~\ref{fig:appendix2}. 

{\it Ferromagnetic properties.} The ferromagnetic properties of the films have been characterized by means of XMCD experiments at 4 K, i.e. well within the FM phase of LSMO. As shown in the inset of Fig.~\ref{fig:appendix1}(c), a field of 3 T is more than enough to magnetically saturate the sample, both in- and out-of-plane. Application of the so-called sum rules~\cite{tholePRL1992, carraPRL1993} should allow for a precise quantitative determination of both spin (m$_s$) and orbital moments (m$_l$) of the element under investigation as long as the intermixing of the $L_3$ and $L_2$ parts of the spectrum is negligible.~\cite{carraPRL1993} Although this is not the case of Mn~[\onlinecite{crocombetteJPCM1996}] a relative comparison is possible. Fig.~\ref{fig:appendix1}(c) shows the calculated spin magnetic moment as function of thickness normalized to that of the thicker sample. As expected, and due to the presence of non-FM layers at the interfaces of LSMO there is a strong reduction of the magnetisation as the film is decreased with a thickness dependency akin to that measured by macroscopic magnetometry methods.~\cite{sandiumengePRL2013} We note that the $9.4$ nm sample, in spite of the presence of a minor Mn$^{2+}$ phase, shows a normalized ms close to that of the thicker $45$ nm film.

{\it Orbital occupation.} X-ray linear dichroism experiments at $\theta=30^o$ and T = $350$ K have been used to characterize the preferential orbital occupation as function of thickness, i.e. XLD$_{\rm OO}  =I^V- I^H = I_{ab}-I_{30^o}^H$, see Fig.~\ref{fig:appendix1}(d). A magnetic field of 3 T has been applied along the beam propagation direction to suppress any spurious FM contribution to the XLD as the magnetisation is saturated and aligned orthogonally to the electric field vector of the incoming radiation (see Fig.~\ref{fig:appendix1}(d)). Similarly to previous reports we observe a change in the sign of the $L_2$ spectral weight as film thickness is reduced thus, highlighting a change from a preferential 3$z^2$-r$^2$ orbital occupation for thinner (t$_{\rm LSMO}  \leq 2.3$ nm) to a x$^2$-y$^2$ one for thicker  (t$_{\rm LSMO}  \geq 3.5$ nm) films, see Fig.~\ref{fig:appendix1}(e). 

Low temperature XLD data was obtained at 4 K. As in the case of the high temperature data, a magnetic field of 3 T was applied along the propagation direction in order to remove the FM contribution to the XLD, hence XLD = XLD$_{\rm OO}$ +XLD$_{\rm AFM}$. Only films with t $_{\rm LSMO}\geq 3.5$ nm show an XLD strongly differing from that measured at 350 K (not shown), likely related to depressed magnetic properties for thinner films.~\cite{sandiumengePRL2013} Selected XLD spectra, normalized to the maximum of the XAS $L_3$ spectral feature for the 8.9 nm, 9.4 nm and 45 nm samples are depicted in Fig.~\ref{fig:appendix2}. As a comparison we also plot the XLD obtained at 350 K for the 9.4 nm sample originating from the orbital occupation (XLD$_{\rm OO}$). The AFM contribution clearly dominates, i.e. at temperatures below the magnetic ordering temperature XLD $\approx$ XLD$_{\rm AFM}$. This is further supported by the similar temperature dependence of both XLD and XMCD for that sample, see Fig.~\ref{fig:setup-XMCD-XLD}(b) in the main text. 

{\it Experimental angular dependence of the XLD.} The angular dependence of the XLD has been measured for $15^o\leq\theta\leq90^o$. As in previous studies~\cite{arutaPRB2009,PesqueraNatComm2012,valenciaJPCM2014,pesqueraPRA2016} the XLD analysis has focused on the $L_2$ spectral region where saturation effects~\cite{nakajimaPRB1999} and possible off-stoichiometry effects~\cite{pesqueraPRA2016} are minimized. Saturation effects, accounting for artificial changes in the absorption due to the angular dependence of the x-ray probing depth, has nonetheless been corrected~\cite{cuiSciRep2014} by using an electron probing depth for TEY of 2.7 nm for LSMO.~\cite{ruosiPRB2014} Fig.~\ref{fig:XLD-exp} in the main text shows the angular dependency of XLD at a given energy across the Mn $L_2$-edge (653.3 eV). In agreement with other studies previously published~\cite{PesqueraNatComm2012,pesqueraPRA2016} selecting another energy and/or integrating the XLD signal leads to similar results.

\section{Theoretical modeling}

The double exchange model H$_{\rm DE}$ is the kinetic energy of a system in which an infinite Hund's coupling between the localized and itinerant carriers forces their spins to be parallel. This model explains the correlation between half-metallic behavior and ferromagnetic order in the optimal doped ($x=1/3$) manganites. It can be written
\begin{equation}
H_{\rm DE}=\sum_{i,j,\alpha,\beta}t_{i,j}^{\alpha,\beta} d_{i\alpha}^{\dagger} d_{j\beta}
\end{equation}
with $t$ the hopping parameter, $i$ and $j$ neighboring Mn sites, and orbital indices representing
either 3z$^2$-r$^2$ or x$^2$-y$^2$, and d the construction/destruction operators. This is a spinless term as the spin of the carriers has been projected on the local spin (S=3/2), which corresponds to the three electrons that occupy the t$_{2g}$ orbitals. The hoppings are anisotropic~\cite{calderonPRB2008} 
\begin{eqnarray}
t_{x(y)}^{x^2-y^2,x^2-y^2}&=&\pm \sqrt{3}t_{x(y)}^{x^2-y^2,3z^2-r^2}=3 t_{x(y)}^{3z^2-r^2,3z^2-r^2}= \nonumber \\
&=& {\frac{3}{4}}t_z^{3z^2-r^2,3z^2-r^2}=t \nonumber \\
t_{z}^{x^2-y^2,x^2-y^2}&=&0
\end{eqnarray}
where the subindices $x, y, z$ refer to the directions in the lattice.
In order to model the surface, it is important to include the long range Coulomb interaction
H$_{\rm Coul}$ between the charges in the system as we expect the charge to redistribute close to the surface. This term is included at the Hartree level~\cite{breyPRB2007}
\begin{equation}
H_{\rm Coul}={\frac{e^2}{\epsilon}} \sum_{i \neq j} \left({\frac{1}{2}}{\frac{\langle n_i\rangle \langle n_j \rangle}{|{\bf R}_i-{\bf R}_j|}}+{\frac{1}{2}}{\frac{Z_i Z_j }{|{\bf R}_i^A-{\bf R}_j^A|}}-{\frac{Z_i \langle n_j \rangle}{|{\bf R}_i-{\bf R}_j^A|}}\right)
\end{equation}
with $R_i$ the position of the Mn ions, $n_i$ the occupation number on the Mn i-site, $eZ_i$ the
charge of the A-cation located at $R_i^A$, and $\epsilon$ the dielectric constant of the material. The relative strength of the Coulomb interaction is given by the parameter $\alpha = e^2/at$, with $a$ the lattice parameter. Here we use $\alpha = 1$. We also consider the spin-orbit interaction H$_{\rm SOC}$ between the e$_g$ orbitals~\cite{fuhrJPCM2010} (see Eq.~\ref{eq:SOI}), which takes place through the t$_{2g}$ orbitals.

\bibliography{spin-orbit}

\end{document}